\documentclass[journal,comml]{IEEEtran}

\usepackage{cite} 
\usepackage{xcolor}
\usepackage{booktabs}
\usepackage{multirow}
\usepackage{graphicx}
\usepackage{amsmath} % Highly recommended for IEEE math environments

\begin{document}
\bstctlcite{IEEEexample:BSTcontrol}
	
		\title{Quantum Meets Statistical-Physical Secrecy: \\A Novel Hybrid Key Distribution Architecture}
	
	\author{Ertugrul~Basar, \textit{Fellow, IEEE}
		 \vspace*{-0.45cm}
		 \thanks{Received 12 May 2026; revised 15 June 2026; accepted
		 	21 June 2026. Date of publication XX XXXX 2026; date of current version
		 	XX XXXX 2026. This work was supported by the TUBITAK under grant no. 124E419. The associate editor coordinating the review of this
		 	letter and approving it for publication was J. Illiano. }
		\thanks{E. Basar is with the Tampere Wireless Research Center, Department of Electrical Engineering, Tampere University, 33720 Tampere, Finland, and holds an adjunct position with the Department of Electrical \& Electronics Engineering, Ko\c{c} University, 34450 Sariyer, Istanbul, T\"{u}rkiye. email: ertugrul.basar@tuni.fi}
		\thanks{Digital Object Identifier 10.1109/LCOMM.2026.XXXXXXX}%
	}
	
	% Optional: Paper headers
%	\markboth{IEEE Communications Letters}%
%	{Basar: Quantum Meets Statistical-Physical Secrecy}

	\maketitle	
	
	\begin{abstract}
	This letter proposes a novel hybrid key distribution architecture that jointly exploits quantum key distribution (QKD) and Kirchhoff-law-Johnson-noise (KLJN) statistical-physical key exchange. In the proposed system, an optical BB84-type QKD link operates in coordination with a parallel wired KLJN link, which is used for secure basis handling and, in selected protocols, additional raw key generation. Three novel KLJN-assisted QKD protocols are introduced to eliminate public basis disclosure messages and bit sifting, extract basis-derived key bits, or generate raw key bits under ideal KLJN assumptions. Analytical expressions for the normalized key rate and absolute throughput are derived by accounting for optical channel penalties, KLJN bandwidth constraints, and synchronization bottlenecks. Numerical results show that the proposed hybrid architecture can improve key generation efficiency and throughput in short-haul infrastructures, including metropolitan area networks (MANs) and data center interconnects.
	
	\end{abstract}
	
	
\vspace*{-0.3cm}

\section{Introduction}

\IEEEPARstart{T}{he} deployment of sixth-generation (6G) networks is expected to impose stringent requirements on secure, resilient, and low-latency connectivity \cite{Porambage_2021}. As metropolitan area networks (MANs), data center interconnects, and mission-critical infrastructures become increasingly dependent on high-capacity links, key distribution mechanisms with information-theoretic security are gaining renewed importance. This trend is further accelerated by the upcoming threat posed by quantum computing to conventional cryptographic techniques.

Quantum key distribution (QKD), particularly the foundational BB84 protocol, provides information-theoretic security based on the laws of quantum mechanics \cite{Bennett_1984}. However, practical QKD systems suffer from optical attenuation, detector imperfections, dark counts, and reconciliation overheads. Moreover, in the widely recognized BB84 protocol, a significant fraction of transmitted quantum states is discarded during basis sifting, since Alice and Bob retain only the events for which their measurement bases coincide \cite{Bennett_1984,Scarani2009}. Although authenticated public discussion does not compromise security directly, public basis disclosure and bit sifting reduce the normalized useful key yield per transmitted optical pulse.

The Kirchhoff-law-Johnson-noise (KLJN) scheme offers an alternative and low-cost key exchange approach by exploiting thermal noise statistics over a wired loop \cite{Kish_2006,Basar_2024}. Under ideal operating conditions, the KLJN scheme can prevent an eavesdropper from distinguishing the two secure mixed-resistance selections. However, the quasi-static requirement that supports KLJN security also limits its native bandwidth, making standalone KLJN throughput strongly constrained by the physical properties of the copper line.

Motivated by the complementary strengths of QKD and KLJN, this letter proposes a novel hybrid QKD-KLJN architecture for short-haul secure key distribution. The main idea is to operate an optical QKD link and a parallel KLJN wired link in a coordinated manner, so that the KLJN subsystem supports secure basis handling of QKD and, in some protocols, contributes additional key bits. Three KLJN-assisted QKD protocols are introduced: Protocol I eliminates public basis discussions and bit sifting; Protocol II additionally extracts basis-derived KLJN key bits; and Protocol III enables one raw key bit per interval under ideal conditions, at the cost of stricter timing requirements.

Under ideal KLJN conditions and appropriate synchronization, the proposed architecture can eliminate public basis announcements and the associated bit sifting process of QKD, improve the normalized key yield, and provide short-range throughput advantages for MANs and data center networks. 

 %\vspace*{-0.1cm}
	\section{The Novel Hybrid QKD-KLJN Architecture}
The proposed architecture consists of two physically distinct but coordinated key distribution links between Alice and Bob: an optical QKD link and a wired KLJN link, as shown in Fig. 1. The optical link (a standard fiber) is responsible for transmitting the polarization-encoded quantum states used in BB84-type operation. In parallel, the KLJN link consists of a bundle of parallel wire pairs, where Alice and Bob connect low or high resistances according to the protocol-dependent basis/resistance mapping. These two links are coordinated by local basis/key mapping, synchronization, and decision units at Alice and Bob, while a conventional authenticated public channel remains available for error correction and privacy amplification for the BB84 system.

\begin{figure*}[!t]
	\begin{center}
		\includegraphics[width=1.54\columnwidth]{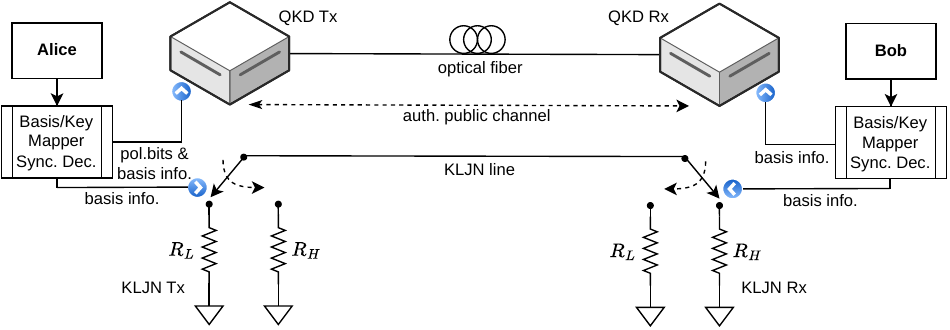}
		\vspace*{-0.3cm}\caption{Proposed hybrid architecture including QKD and KLJN terminals/links and related basis/key mapper, synchronization, and decision units.}\vspace*{-0.5cm}
		\label{fig:NoiseMod}
	\end{center}
\end{figure*} 

The role of the KLJN subsystem depends on the selected protocol, which will be covered in Section III. The proposed architecture can operate in two timing modes. In the real-time gated mode, each KLJN decision is directly synchronized with one optical transmission interval, so that the overall system rate is limited by the slower of the optical repetition rate and the KLJN decision-bit rate. This mode is particularly relevant for Protocol III since it reveals the selected QKD bases. In contrast, Protocols I and II can also support an asynchronous buffered mode: the KLJN subsystem continuously generates secure basis control bits in the background, which are stored in local secure buffers at Alice and Bob. Once sufficient buffered bits are available, the QKD subsystem can operate at its native repetition rate. Although the long-term average throughput remains limited by the KLJN replenishment rate, the buffered operation allows high-speed optical hardware to be used without continuous down-clocking.

\vspace*{-0.15cm}
\section{Proposed secure protocols}

In this section, we introduce three novel protocols for our hybrid architecture, as shown in Table I through examples.

In Protocol I, we follow the same procedures of BB84 protocol; however, the public discussion for selected bases is eliminated thanks to the specific operation of the KLJN subsystem in parallel through a wired line. Here, the KLJN subsystem is used to securely share the basis information only and the key is generated according to quantum measurements as in classical BB84 protocol. Specifically, Alice and Bob follow the following cross mapping to select their resistances:
\begin{align}
	\text{Alice Basis } +/\times &\rightarrow \text{Alice Res. } R_L/R_H \nonumber \\ 
	\text{Bob Basis } +/\times &\rightarrow \text{Bob Res. } R_H/R_L.
\end{align}
Since different bases are encoded into different resistances at Alice and Bob, when Alice and Bob select the same basis, an intermediate noise level will be observed on the KLJN line. That is, $+/+$ and $\times/\times$ bases selections of Alice/Bob would correspond $R_L/R_H$ and $R_H/R_L$, which will be indistinguishable for Eve under ideal KLJN operating conditions. Here, ideal KLJN conditions refer to the quasi-static KLJN model in which the two secure mixed-resistance states produce indistinguishable voltage/current observations for Eve. This assumes matched effective noise temperatures, negligible or compensated line nonidealities, and the absence or successful detection/mitigation of active attacks, such as transient or current-injection attacks. A complete security proof under practical nonidealities is beyond the scope of this short letter.

On the other hand, different bases selection of Alice and Bob, $+/\times$ or $\times/+$, correspond to low $(R_L/R_L)$ or high $(R_H/R_H)$ variances on the line, respectively, which are already marked as insecure by the KLJN system. An eavesdropper measuring the KLJN line can understand the cases where Alice’s and Bob’s bases overlap; however, it cannot determine their specific selections. Therefore, Eve cannot learn the actual common basis of the retained QKD events, and passive KLJN observation does not alter the standard BB84 measurement-disturbance mechanism or the subsequent error rate estimation. This information is obvious to Alice and Bob since they know their own resistance and can understand and find their partner’s resistance/basis. Since Alice and Bob can also observe low and high variances through the KLJN line, they understand that in these cases their quantum bases do not overlap; as a result, they can perform bit sifting instantly and discard these bits without any public discussion. We note that QKD-derived bit is retained if the corresponding optical pulse also produces a valid detection event when an intermediate noise level is observed on the KLJN line. 

To sum up, Protocol I generates the same sifted secure key as the BB84 protocol; however, it performs this without any public discussion for bases and bit sifting. On the other hand, due to errors on the quantum channel, the two terminals  still need to perform classical post-processing (with a lower overhead) over a public channel to generate a secure key. 

Protocol II follows the exact bit mapping as Protocol I; however, it generates additional key bits through the KLJN subsystem. Since the contributions of Alice and Bob to the intermediate noise level on the wire cannot be distinguished under ideal KLJN conditions, with their specific bit selection, one additional secure key bit can be generated, as follows 
\begin{align}
	\text{Bases (Alice/Bob)} = (+/ +) &\rightarrow \text{Res. } (R_L/R_H) \rightarrow \text{Bit } 0 \nonumber \\
	\text{Bases (Alice/Bob)} = (\times/\times) &\rightarrow \text{Res. } (R_H/R_L) \rightarrow \text{Bit } 1.  \nonumber 
\end{align}  
As seen from Table I, Alice and Bob's bases overlap for nine occasions, and nine additional key bits are created by Protocol II, considering Alice and Bob's resistor selections through the KLJN line. In other words, the common basis label is used for KLJN-derived key generation, while the QKD-derived bits come from the polarization bit values. Under random and unbiased BB84 bit/basis selection, these contributions are statistically independent and Eve cannot distinguish ($+$/$+$) and ($\times$/$\times$) cases. For Alice and Bob, due to knowledge of their own resistor, understanding their partner's resistor/basis is straightforward, and total QKD+KLJN bit sifting can be performed without any public discussion or revealing their bases. 

\begin{table*}[t]
	\centering
	\caption{Secure key generation steps for Protocols I, II, and III }
	\vspace*{-0.2cm}
	\label{tab:qkd_kljn_protocols}
	% Optional: If the table is too wide for the margins, uncomment the resizebox line below
	% \resizebox{\textwidth}{!}{%
		\begin{tabular}{ll *{14}{c}}
			\toprule
			& \textbf{Alice's Basis} & $+$ & $+$ & $\times$ & $+$ & $\times$ & $\times$ & $\times$ & $+$ & $\times$ & $+$ & $+$ & $\times$ & $\times$ & $+$ \\
			& \textbf{Alice's Bit} & 1 & 0 & 1 & 1 & 0 & 0 & 1 & 1 & 0 & 0 & 1 & 1 & 1 & 0 \\
			& \textbf{Alice's Polarization} & $\uparrow$ & $\rightarrow$ & $\nearrow$ & $\uparrow$ & $\searrow$ & $\searrow$ & $\nearrow$ & $\uparrow$ & $\searrow$ & $\rightarrow$ & $\uparrow$ & $\nearrow$ & $\nearrow$ & $\rightarrow$ \\
			\midrule
			& \textbf{Bob's Basis} & $+$ & $\times$ & $+$ & $+$ & $\times$ & $\times$ & $+$ & $+$ & $\times$ & $+$ & $\times$ & $\times$ & $+$ & $+$ \\
			& \textbf{Bob's Polarization} & $\uparrow$ & $\nearrow$ & $\rightarrow$ & $\uparrow$ & $\searrow$ & $\searrow$ & $\uparrow$ & $\uparrow$ & $\searrow$ & $\rightarrow$ & $\nearrow$ & $\nearrow$ & $\rightarrow$ & $\rightarrow$ \\
			& \textbf{Bob's Bit} & 1 & 1 & 0 & 1 & 0 & 0 & 1 & 1 & 0 & 0 & 1 & 1 & 0 & 0 \\
			\midrule
			\multirow{3}{*}{\textbf{I}} 
			& \textbf{Alice's Resistor} & $R_L$ & $R_L$ & $R_H$ & $R_L$ & $R_H$ & $R_H$ & $R_H$ & $R_L$ & $R_H$ & $R_L$ & $R_L$ & $R_H$ & $R_H$ & $R_L$ \\
			& \textbf{Bob's Resistor} & $R_H$ & $R_L$ & $R_H$ & $R_H$ & $R_L$ & $R_L$ & $R_H$ & $R_H$ & $R_L$ & $R_H$ & $R_L$ & $R_L$ & $R_H$ & $R_H$ \\
			& \textbf{Generated/Sifted Key} & \textcolor{red}{1} & -- & -- & \textcolor{red}{1} & \textcolor{red}{0} & \textcolor{red}{0} & -- & \textcolor{red}{1} & \textcolor{red}{0} & \textcolor{red}{0} & -- & \textcolor{red}{1} & -- & \textcolor{red}{0} \\
			\midrule
			\multirow{3}{*}{\textbf{II}} 
			& \textbf{Alice's Resistor} & $R_L$ & $R_L$ & $R_H$ & $R_L$ & $R_H$ & $R_H$ & $R_H$ & $R_L$ & $R_H$ & $R_L$ & $R_L$ & $R_H$ & $R_H$ & $R_L$ \\
			& \textbf{Bob's Resistor} & $R_H$ & $R_L$ & $R_H$ & $R_H$ & $R_L$ & $R_L$ & $R_H$ & $R_H$ & $R_L$ & $R_H$ & $R_L$ & $R_L$ & $R_H$ & $R_H$ \\
			& \textbf{Generated/Sifted Key} & \textcolor{red}{1} \textcolor{green}{0} & -- & -- &  \textcolor{red}{1} \textcolor{green}{0} &  \textcolor{red}{0} \textcolor{green}{1} &  \textcolor{red}{0} \textcolor{green}{1} & -- &  \textcolor{red}{1} \textcolor{green}{0} &  \textcolor{red}{0} \textcolor{green}{1} &  \textcolor{red}{0} \textcolor{green}{0} & -- &  \textcolor{red}{1} \textcolor{green}{1} & -- &  \textcolor{red}{0} \textcolor{green}{0} \\
			\midrule
			\multirow{3}{*}{\textbf{III}} 
			& \textbf{Alice's Resistor} & $R_L$ & $R_L$ & $R_H$ & $R_L$ & $R_H$ & $R_H$ & $R_H$ & $R_L$ & $R_H$ & $R_L$ & $R_L$ & $R_H$ & $R_H$ & $R_L$ \\
			& \textbf{Bob's Resistor} & $R_L$ & $R_H$ & $R_L$ & $R_L$ & $R_H$ & $R_H$ & $R_L$ & $R_L$ & $R_H$ & $R_L$ & $R_H$ & $R_H$ & $R_L$ & $R_L$ \\
			& \textbf{Generated/Sifted Key} & \textcolor{red}{1} & \textcolor{green}{0}  & \textcolor{green}{1}  & \textcolor{red}{1} & \textcolor{red}{0} & \textcolor{red}{0} & \textcolor{green}{1}  &  \textcolor{red}{1} & \textcolor{red}{0} & \textcolor{red}{0} & \textcolor{green}{0} & \textcolor{red}{1} & \textcolor{green}{1} & \textcolor{red}{0} \\
			\bottomrule
		\end{tabular}
		\par
		\vspace{1ex}
		{\raggedright \scriptsize  \textbf{Bases:} $+$ (Rectilinear), $\times$ (Diagonal). \textbf{Polarizations:} $\uparrow$ (Vertical, Bit 1), $\rightarrow$ (Horizontal, Bit 0), $\nearrow$ ($45^\circ$, Bit 1), $\searrow$ ($-45^\circ$, Bit 0). \textbf{Colors:} Red (QKD sifted bits), Green (KLJN basis-derived bits).\par} \vspace*{-0.4cm}
	\end{table*}

Protocol III differs from Protocols I and II in operation and basis/resistor mapping.
Specifically, in Protocol III, Alice and Bob consider the same mapping given below:
\begin{align}
	\text{Basis } +/\times  &\rightarrow \text{Res. } R_L/R_H.
\end{align}
%With the above resistor/basis mapping, we have the following four different scenarios for the KLJN subsystem: 
%\begin{enumerate}
%	\item Alice Basis $+$ and Bob Basis $+$ $\rightarrow$ Alice Res. $R_L$ and Bob Res. $R_L$ $\rightarrow$ Noise Level: Low %
%	\item Alice Basis $+$ and Bob Basis $\times$ $\rightarrow$ Alice Res. $R_L$ and Bob Res. $R_H$ $\rightarrow$ Noise Level: Intermediate %
%	\item Alice Basis $\times$ and Bob Basis $+$ $\rightarrow$ Alice Res. $R_H$ and Bob Res. $R_L$ $\rightarrow$ Noise Level: Intermediate %
%	\item Alice Basis $\times$ and Bob Basis $\times$ $\rightarrow$ Alice Res. $R_H$ and Bob Res. $R_H$ $\rightarrow$ Noise Level: High.
%\end{enumerate}
%In cases 1 and 4, we use QKD measurements for secure key generation since Alice and Bob have the same bases. However, their basis information is no longer secret due to the selection of the same resistor/basis mapping at Alice and Bob; that is, if they overlap on the rectilinear ($+$) basis, the KLJN noise level will be low, and if they overlap on the diagonal ($\times$) basis, it will be high. On the other hand, in cases 2 and 3, we use KLJN measurements only for secure key generation since an intermediate noise level is observed on the KLJN line. Similar to Protocol II, the order of Alice and Bob's resistors is used to generate the secure key bit:
%\begin{align}
%	\text{Bases (Alice/Bob)} = (+/ \times) &\rightarrow \text{Res. } (R_L/ R_H) \rightarrow \text{Bit } 0 \nonumber \\
%	\text{Bases (Alice/Bob)} = (\times/ +) &\rightarrow \text{Res. } (R_H/ R_L) \rightarrow \text{Bit } 1 \nonumber 
%\end{align}
With the above basis/resistor mapping, matched bases produce publicly distinguishable low/high KLJN levels, namely $(+/+ \rightarrow R_L/R_L)$ and $(\times/\times \rightarrow R_H/R_H)$, while mismatched bases produce the intermediate KLJN states, $(+/\times \rightarrow R_L/R_H)$ and $(\times/+ \rightarrow R_H/R_L)$. Therefore, QKD measurements are used for key generation in the matched-basis cases, whereas KLJN measurements are used in the mismatched-basis cases. Similar to Protocol II, the order of Alice's and Bob's resistors determines the KLJN-derived bit: $(+/\times \rightarrow R_L/R_H \rightarrow 0)$ and $(\times/+ \rightarrow R_H/R_L \rightarrow 1)$.

As seen from Table I, Protocol III can generate a secure bit in all transmission intervals (in the ideal lossless limit) by alternating between QKD- and KLJN-generated key bits. This ensures that Protocol III does not need bit sifting, but the price paid is for revealing the common basis information in the QKD system through the low and high noise variance levels in the KLJN system. This necessitates firing optical pulses before the KLJN operation in a gated synchronous mode, so that BB84 security is not compromised. Thus, Protocol III should be interpreted as a timing-constrained ideal variant, where the optical measurement is completed before the KLJN-based basis disclosure; a full security analysis of joint quantum-KLJN attacks is left for future work.

\vspace*{-0.3cm}
\section{System Model and Analyses}

To evaluate the performance of the proposed hybrid QKD-KLJN architecture against the standard BB84 protocol, we develop a rigorous system model. Our model separates the physical boundaries of the copper infrastructure used for KLJN, the cryptographic penalties of the quantum optical channel, and the hardware synchronization bottlenecks of the overall system. Since the main objective is to evaluate the proposed QKD-KLJN protocols, we adopt a simplified asymptotic normalized key rate model.

For simplicity, we consider an idealized, single-photon-equivalent BB84 baseline without decoy states. The physical transmission over the optical fiber is modeled using a standard weak coherent pulse (WCP) channel with Poisson photon-number statistics. Following the framework established by \cite{Ma2005}, the overall system transmittance $\eta_{sys}$ is defined as a function of the detector efficiency $\eta_D$, the fiber attenuation $\alpha$ (in dB/km), and the transmission distance $L$ (in km): $\eta_{sys} = \eta_D 10^{-\frac{\alpha L}{10}}$. Accordingly, the overall expected photon gain and the quantum bit error rate are given respectively as $Q_\mu = 1 - e^{-\mu  \eta_{sys}} + p_d $ and $E_\mu = (e_{opt}  (1 - e^{-\mu  \eta_{sys}}) + 0.5  p_d) / Q_\mu$,
%\begin{align}
%	Q_\mu &= 1 - e^{-\mu  \eta_{sys}} + p_d  \nonumber \\
%		E_\mu &= (e_{opt}  (1 - e^{-\mu  \eta_{sys}}) + 0.5  p_d) / Q_\mu
%\end{align}
where $\mu$ is the mean photon number per pulse, $e_{opt}$ represents the optical misalignment error, and $p_d$ is the detector dark count probability per pulse. 
% In other words, $Q_\mu$ stands for the probability that an emitted signal (in this case, a Poisson-distributed WCP) will lead to a detection event at Bob's side. The term $0.5  p_d$ models the randomized error contribution of detector dark counts \cite{Ma2005}.

To distill a secure key, the raw optical fraction is penalized by a dynamic overhead ($\gamma$). Motivated by GLLP-style post-processing penalties for imperfect-device QKD \cite{GLLP2004}, we  use a simplified asymptotic penalty factor for both practical error correction inefficiency ($f$) and the privacy amplification required to erase potential collective attacks by an eavesdropper. This penalty term is bounded by unity (representing 100\% overhead) as $	\gamma = \min\big(1, (f + 1) h(E_\mu)\big)$, where $h(E_\mu)$ is the binary entropy function.
%where $h(E_\mu) = -E_\mu \log_2(E_\mu) - (1 - E_\mu) \log_2(1 - E_\mu)$ is the binary entropy function.

To preserve the security of the KLJN scheme, the noise bandwidth must be constrained by the physical limits of the wire line \cite{Kish_2016}. In practical cables, distributed parasitic capacitance introduces an RC-related nonideality that can lead to information leakage and impose an additional effective bandwidth constraint. This leakage can be mitigated using capacitor-killer or capacitance-compensation techniques \cite{Chen_2015}. Therefore, assuming ideal capacitance compensation, we focus on the quasi-static wave limit ($B_W$) as the dominant bandwidth constraint for short-haul links. The quasi-static condition requires the cable to behave as a lumped electrical system, thereby suppressing wave-propagation effects, standing waves, and reflections. Equivalently, the transmission distance ($L$) must be much smaller than the signal wavelength ($\lambda$), i.e., $L \ll \lambda$. As a practical safety margin, we require the highest noise frequency to be at least ten times lower than the lowest standing-wave frequency of the line, $f_1 = v/(2L)$, where $v$ is the signal velocity in copper. Hence, the allowable noise bandwidth is bounded as $B_W = v/(20L)$. By bounding the analog (external) noise generator to this maximum safe frequency, the effective ADC sampling frequency ($f_{s}$) is dictated by the Nyquist theorem applied directly to the wave limit: $f_s=2 B_W$.

 Unlike standard digital communication where a single clock cycle yields a data bit, the nature of the KLJN protocol dictates that bits (noise levels) cannot be detected instantaneously. Instead, Alice and Bob must collect  $N$ discrete voltage and/or current samples to distinguish different noise levels reliably. Fortunately, alternating voltage and current measurements can be used to achieve very low bit error probability (close to $10^{-5}$) with small $N$ values and such residual KLJN errors are neglected \cite{Basar_2023}. To compensate for this sampling overhead and overcome the limited bandwidth of a single copper channel, our architecture employs spatial multiplexing by bundling multiple parallel wire pairs ($N_{pairs}$) within a single cable. By dividing the available sampling frequency by the required samples per bit and aggregating the parallel streams, the total KLJN decision-bit rate $R_{KLJN}$ (in bits per second, bps) is formulated as: $	R_{KLJN} = N_{pairs}f_s/N$.

Assuming that hybrid architecture employs a one-to-one gated synchronization, where each generated KLJN bit corresponds exactly one optical pulse, the aggregated decision-bit rate $R_{KLJN}$ physically translates to an equivalent optical trigger demand in Hz. Therefore, the effective system trigger frequency $f_{sys}$ is bounded by the physical limitations of the slowest hardware component as $f_{sys} = \min(f_{QKD}, R_{KLJN})$. Here, $f_{QKD}$ is the native maximum repetition rate of the optical source. It is worth noting that due to the high frequency of the optical source, $f_{sys}$ is mainly dominated by $R_{KLJN}$, which drops dramatically by increasing distance.

The normalized key rate (in secure bits/pulse) defines the theoretical efficiency of a single transmission cycle. Standard BB84 and the unassisted Protocol I are strictly bound by the optical penalty:
\begin{equation}
	R_{BB84} = R_I = 0.5  Q_\mu  (1 - \gamma).
\end{equation}
Conversely, the proposed Protocols II and III generate additional key bits through the KLJN subsystem, resulting in a KLJN-derived contribution under the idealized KLJN model, valid KLJN decisions, and unbiased basis selection:
\begin{equation}
	R_{II, III} = 0.5  Q_\mu (1 - \gamma) + 0.5.
\end{equation}
It is worth noting that our protocols still suffer from the penalty of $(1-\gamma)$ due to the classical post-processing steps required for BB84-generated bits to ensure unconditional security.

The absolute throughput (key rate, in bps) applies the hardware constraints to the normalized rates. Standard BB84 operates completely unthrottled over the optical link, whereas the hybrid protocols are bounded by the system trigger frequency and severely throttled by $R_{KLJN}$:
\begin{gather}
	T_{BB84} = R_{BB84} f_{QKD} \nonumber \\
		T_I = R_I  f_{sys}  \quad \quad 	T_{II, III} = R_{II, III}  f_{sys}.
		\label{eq:T}
\end{gather}

The absolute throughput formulation in \eqref{eq:T} assumes a strict, real-time one-to-one gated synchronization scheme where the optical source is continuously throttled by the limits of the classical line. While this real-time operation highlights the fundamental physical bottleneck of the system, it represents a worst case temporal scenario. This severe throttling can be circumvented by shifting the architecture from real-time generation to an asynchronous burst-mode operation.

Since the KLJN wire channel and the QKD optical fiber are physically independent transmission media, their execution does not need to be strictly simultaneous. In an asynchronous buffering scheme, the KLJN terminals operate continuously in the background at their maximum physically allowed rate ($R_{KLJN}$). The securely generated KLJN bits, which govern the basis choices for our protocols, are accumulated in a secure local memory buffer at both Alice's and Bob's terminals. During this classical accumulation phase, the optical QKD system remains inactive. Since Protocol III reveals the selected bases, it could compromise standard BB84 security; therefore, it must be used in gated synchronous (real-time) mode. 

Once a sufficient block of classical bits is populated in the buffer, the QKD system can be activated. Because the basis choices are pre-established and securely stored, the optical transmitter can sequentially consume the buffered bits and fire at its native, unthrottled maximum repetition rate ($f_{QKD}$). During this optical transmission phase, the transient (burst) throughput of the hybrid protocols escapes the KLJN throttling effect, achieving parity with unassisted QKD systems: $T_{burst, x} = R_{x} f_{QKD}, \, \text{for } x \in \{I, II\}$. Under this buffering paradigm, Protocols I and II can fully exploit standard high-speed telecom lasers and detectors without requiring down-clocking of the optical hardware. It must be acknowledged that the long-term average throughput over an extended operational period remains fundamentally bounded by the classical channel's capacity to replenish the basis buffer and follows  \eqref{eq:T}. However, this asynchronous buffering strategy offers a deployment advantage for the integration of thermodynamic security into existing high-speed QKD infrastructures without compromising their burst transmission capabilities.

\begin{table}[!t]
	\centering
	\caption{Computer Simulation Parameters}
	\label{tab:sim_params}
	\begin{tabular}{@{}llc@{}}
		\toprule
		\textbf{Symbol} & \textbf{Parameter Description} & \textbf{Value} \\ \midrule
	%	\multicolumn{3}{c}{\textit{Quantum Optical Domain (WCP)}} \\
		$\alpha$ & Fiber attenuation (1550 nm) & $0.2$ dB/km \\
		$f_{QKD}$ & QKD laser repetition rate & $10$ MHz \\
		$\mu$ & Mean photon number per pulse & $0.1$ \\
		$\eta_D$ & Detector efficiency & $0.1$ \\
		$p_d$ & Dark count probability per pulse & $10^{-5}$ \\
		$e_{opt}$ & Optical misalignment error & $0.015$ \\
		$f$ & Error correction inefficiency factor & $1.15$ \\ 
	%	\multicolumn{3}{c}{\textit{Classical Hardware Domain (KLJN)}} \\
		$v$ & Signal velocity in copper & $2 \times 10^5$ km/s \\
	%	$c_0$ & Cable parasitic capacitance & $50$ nF/km \\
%		$R_{eq}$ & Effective KLJN source resistance & $1000$ $\Omega$ \\
		$N_{pairs}$ & Spatial multiplexing pairs & $500, 1000$ \\
		$N$ & Gaussian noise samples per bit & $50$ \\ \bottomrule
	\end{tabular}
	\vspace*{-0.4cm}
\end{table}

%	\vspace*{-0.25cm}
\section{Numerical Results}
The  used computer simulation parameters are selected according to realistic WCP implementations operating in the telecom C-band (1550 nm) and standard copper wire limits. The complete set of parameters used for the numerical evaluations is provided in Table \ref{tab:sim_params}. For optical fiber and detector characteristics, standard telecom fiber QKD demonstrations \cite{Gobby2004}, WCP/decoy-state modeling conventions \cite{Ma2005}, and practical QKD reviews \cite{Scarani2009} are considered. A relatively conservative QKD laser repetition rate ($f_{QKD}$) is selected to accommodate the one-to-one gated synchronization architecture of the proposed protocols. Similarly, representative values from the literature are selected for KLJN \cite{Kish_2016,Chen_2015,Basar_2023}. The $10$ km range should be interpreted as an idealized short-haul benchmark in which the corresponding wire resistance and related leakage mechanisms are neglected.

\begin{figure}[!t]
	\begin{center}
		\includegraphics[width=0.65\columnwidth]{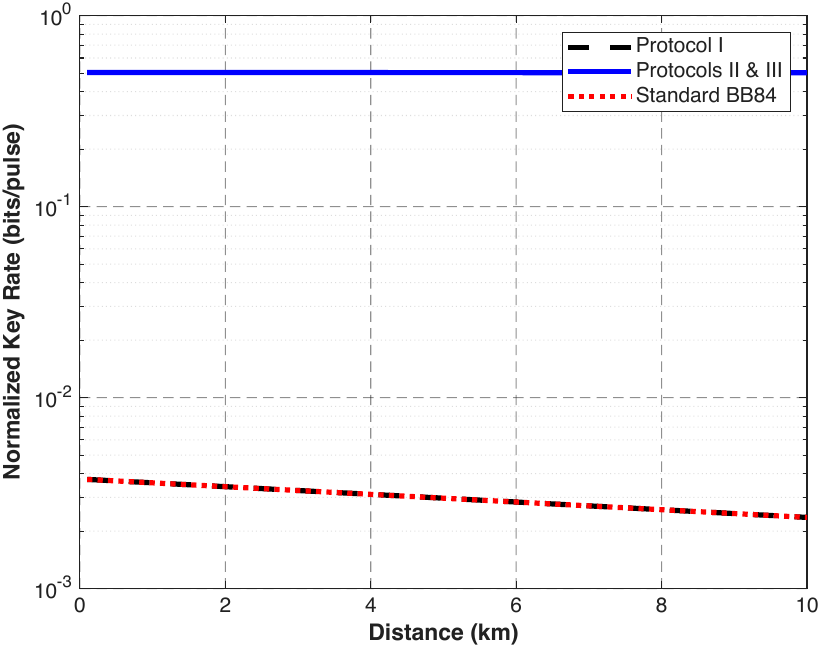}
		\vspace*{-0.3cm}\caption{Normalized key rate for Protocols I, II, and III and the BB84 protocol.}\vspace*{-0.5cm}
	\end{center}
\end{figure}

Fig. 2 illustrates the normalized key rate, which defines the system's efficiency per transmission cycle. As shown, the standard BB84 protocol suffers from optical attenuation, yielding a diminishing fraction of a bit per emitted pulse. Conversely, the hybrid protocols (II and III) maintain a relatively flat and highly efficient normalized rate strictly bounded above $0.5$ bits/pulse. This massive efficiency gap is a direct result of the gated synchronization architecture. By anchoring each optical pulse to a classical KLJN bit exchange, the system guarantees a baseline yield of $0.5$ secure bits per cycle, independent of optical fiber losses. Consequently, the hybrid architecture prevents wasted transmission cycles, ensuring that even if an optical pulse is lost to attenuation, the cycle still  generates secure bits via the classical channel.

\begin{figure}[!t]
	\begin{center}
		\includegraphics[width=0.65\columnwidth]{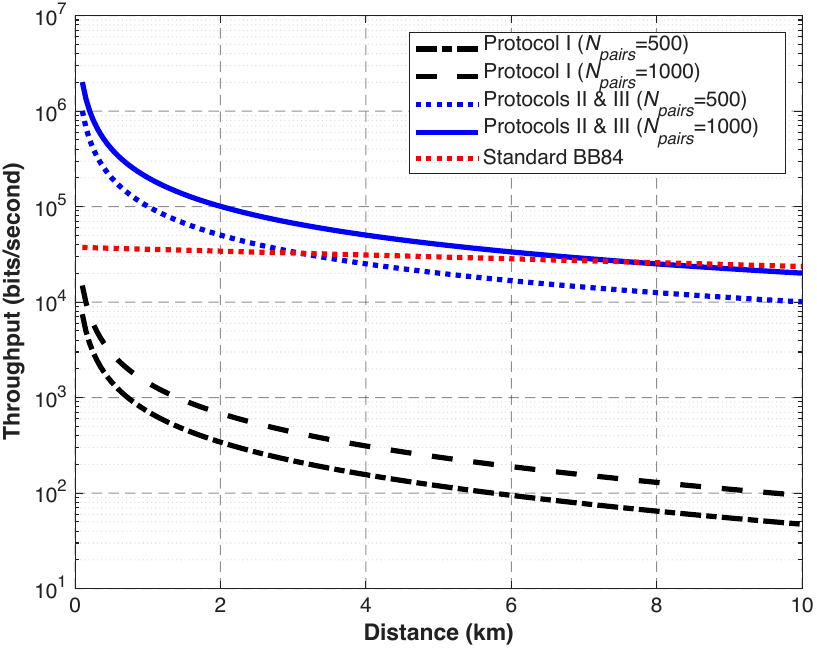}
		\vspace*{-0.3cm}\caption{Absolute throughput for Protocols I, II, and III and the BB84 protocol.}\vspace*{-0.5cm}
	\end{center}
\end{figure} 

Fig. 3 presents the secure throughput of the proposed protocols as a function of transmission distance over a 10 km MAN range. Unlike the normalized key rate, which isolates the efficiency of a single cycle, the absolute throughput is strictly constrained by the synchronization bottleneck ($f_{sys}$) for our protocols. Since the KLJN decision-bit rate scales linearly with the number of parallel wire pairs, we consider two different $N_{pairs}$ values. Here, the short-haul advantage of Protocols II and III should be interpreted as conditional on the availability of spatially multiplexed copper infrastructure.

At distances below $3$~km, Protocols II and III achieve significant short-haul advantage, easily outperforming standard BB84 thanks to the spatial multiplexing of the parallel KLJN lines. While providing unconditionally secure basis reconciliation, Protocol I falls far behind, as it is throttled by the classical line without gaining the classical bit yield. As distance increases, the quasi-static wave limit starts to throttle the classical key generation rate. Because the absolute throughput of the hybrid system is fundamentally bounded by this classical bottleneck, it decays inversely with distance and crosses below the unthrottled BB84 curve at roughly $3$ km and $7.5$ km for  $N_{pairs}=500$ and $1000$, respectively. Despite this crossover, the hybrid architecture maintains a practical short-range throughput on the order of $10^4$ bps over the considered $10$ km range. This confirms that while the physical limits of the copper line restrict the hybrid architecture from long-haul deployment, it provides a potentially advantageous hybrid solution for short-haul networks, data center interconnects, and intra-city MANs.

%\vspace*{-0.30cm}
\section{Conclusions and Future Directions}
This preliminary letter has introduced a novel hybrid QKD-KLJN key distribution architecture in which quantum optical transmission is complemented by statistical-physical secrecy over a parallel wired KLJN line. Three protocols have been proposed to exploit the KLJN subsystem for secure basis handling and additional key generation. Under ideal KLJN assumptions, the proposed architecture can eliminate public basis reconciliation, improve normalized key generation efficiency, and provide noticeable short-range throughput gains compared with conventional BB84 operation. 

While the current architecture utilizes a standard KLJN configuration, its more advanced variants can be considered to neutralize wire resistance attacks. This would enable much longer transmission distances without requiring impractically thick cable diameters or privacy amplification penalties. Furthermore, more advanced KLJN-aided protocols can be developed in the future to completely eliminate public discussion channels of BB84 systems. Finally, a more sophisticated threat model can be established in future studies, specifying Eve's access to the optical channel, KLJN line, timing information, and possible active attacks.

%\begin{acks}
%Insert the Acknowledgment text here.
%\end{acks}


%\bibliographystyle{iet}
%\bibliography{iet-ell}
%
% once the .bbl file has been generated then place the text in your article.

%\vspace*{-0.3cm}
\bibliographystyle{IEEEtran}
\bibliography{IEEEabrv, bib_2026}

\begin{thebibliography}{10}
\providecommand{\url}[1]{#1}
\csname url@samestyle\endcsname
\providecommand{\newblock}{\relax}
\providecommand{\bibinfo}[2]{#2}
\providecommand{\BIBentrySTDinterwordspacing}{\spaceskip=0pt\relax}
\providecommand{\BIBentryALTinterwordstretchfactor}{4}
\providecommand{\BIBentryALTinterwordspacing}{\spaceskip=\fontdimen2\font plus
\BIBentryALTinterwordstretchfactor\fontdimen3\font minus
  \fontdimen4\font\relax}
\providecommand{\BIBforeignlanguage}[2]{{%
\expandafter\ifx\csname l@#1\endcsname\relax
\typeout{** WARNING: IEEEtran.bst: No hyphenation pattern has been}%
\typeout{** loaded for the language `#1'. Using the pattern for}%
\typeout{** the default language instead.}%
\else
\language=\csname l@#1\endcsname
\fi
#2}}
\providecommand{\BIBdecl}{\relax}
\BIBdecl

\bibitem{Porambage_2021}
P.~Porambage \emph{et~al.}, ``The roadmap to 6{G} security and privacy,''
  \emph{IEEE Open J. Commun. Soc.}, vol.~2, pp. 1094--1122, 2021.

\bibitem{Bennett_1984}
C.~H. Bennett and G.~Brassard, ``Quantum cryptography: {P}ublic key
  distribution and coin tossing,'' in \emph{Proc. IEEE Int. Conf. Comput.,
  Syst. Signal Process.}, Bangalore, India, 1984, pp. 175--179.

\bibitem{Scarani2009}
V.~Scarani \emph{et~al.}, ``The security of practical quantum key
  distribution,'' \emph{Rev. Mod. Phys.}, vol.~81, no.~3, pp. 1301--1350, 2009.

\bibitem{Kish_2006}
L.~B. Kish, ``Totally secure classical communication utilizing {J}ohnson
  (-like) noise and {K}irchoff's law,'' \emph{Phys. Lett. A}, vol. 352, pp.
  178--182, 2006.

\bibitem{Basar_2024}
E.~Basar, ``{K}irchhoff meets {J}ohnson: {I}n pursuit of unconditionally secure
  communication,'' \emph{Eng. Rep.}, vol.~6, no.~10, p. e12958, 2024.

\bibitem{Ma2005}
X.~Ma \emph{et~al.}, ``Practical decoy state for quantum key distribution,''
  \emph{Phys. Rev. A}, vol.~72, p. 012326, 2005.

\bibitem{GLLP2004}
D.~Gottesman \emph{et~al.}, ``Security of quantum key distribution with
  imperfect devices,'' \emph{Quantum Inf. Comput.}, vol.~4, no.~5, pp.
  325--360, 2004.

\bibitem{Kish_2016}
L.~B. Kish, \emph{The Kish Cypher: The Story of KLJN for Unconditional
  Security}.\hskip 1em plus 0.5em minus 0.4em\relax World Scientific Publishing
  Co. Pte. Ltd., 2017.

\bibitem{Chen_2015}
H.-P. Chen \emph{et~al.}, ``Cable capacitance attack against the {KLJN} secure
  key exchange,'' \emph{Information}, vol.~6, no.~4, pp. 719--732, 2015.

\bibitem{Basar_2023}
E.~Basar, ``Communication by means of thermal noise: {T}oward networks with
  extremely low power consumption,'' \emph{IEEE Trans. Commun.}, vol.~71,
  no.~2, pp. 688--699, Feb. 2023.

\bibitem{Gobby2004}
C.~Gobby, Z.~L. Yuan, and A.~J. Shields, ``Quantum key distribution over 122 km
  of standard telecom fiber,'' \emph{Appl. Phys. Lett.}, vol.~84, no.~19, pp.
  3762--3764, 2004.

\end{thebibliography}

\end{document}